\documentclass[aps,reprint,amsmath,amssymb,showpacs,showkeys]{revtex4-1}

\usepackage{graphicx}  
\usepackage{natbib}  
\usepackage{amsmath}  
\usepackage{hyperref}  
\usepackage[top=1.5in, bottom=1.5in, left=1in, right=1in]{geometry}
\usepackage{caption}
\usepackage{amsfonts}
\usepackage{xcolor}
\usepackage{booktabs}
\usepackage{multirow}

\begin{document}\sloppy  

\title{Classifying the Equation of State from Rotating Core Collapse
  Gravitational Waves with Deep Learning}

\author{Matthew C. Edwards} \affiliation{Department of Statistics,
  University of Auckland, Auckland, New Zealand}

\begin{abstract}

  In this paper, we seek to answer the question ``\textit{given a
    rotating core collapse gravitational wave signal, can we determine
    its nuclear equation of state?}''.  To answer this question, we
  employ deep convolutional neural networks to learn visual and
  temporal patterns embedded within rotating core collapse
  gravitational wave (GW) signals in order to predict the nuclear
  equation of state (EOS). Using the 1824 rotating core collapse GW
  simulations by \citet{richers:2017}, which has 18 different nuclear
  EOS, we consider this to be a classic multi-class image
  classification and sequence classification problem.  We attain up to
  72\% correct classifications in the test set, and if we consider the
  ``top 5'' most probable labels, this increases to up to 97\%,
  demonstrating that there is a moderate and measurable dependence of
  the rotating core collapse GW signal on the nuclear EOS.
  
\end{abstract}


\maketitle

\section{Introduction}\label{sec:intro}

To date, gravitational waves (GWs) from stellar core collapse have not
been directly observed by the network of terrestrial detectors,
Advanced LIGO and Advanced Virgo \citep{PhysRevD.101.084002}.
However, they are a promising source \citep{gossan:2016}, and we could
learn a great deal about the dynamics of the core collapse, and the
shock revival mechanism that leads to explosion \citep{Kuroda_2017}.
It may even be possible to constrain the nuclear equation of state
(EOS).  

The death of massive stars (of at least 10 $M_{\odot}$ at ZAMS) begins
when the star exhausts its thermonuclear fuel through fusion, leaving
an iron core that is supported by the pressure of relativistic
degenerate electrons.  Once the core reaches the Chandrasekhar limit,
photodissociation of heavy nuclei initiates the collapse, and a
reduction of electron degeneracy pressure accelerates it.  The core
compresses, increasing in density, and squeezing protons and electrons
together to create neutrons and neutrinos via electron-capture.  The
strong nuclear force halts the collapse by a stiffening of the nuclear
EOS, which causes the inner core to rebound (or bounce), creating a
shock wave that blasts into the in-falling outer core.  The shock wave
on its own is not strong enough to generate a supernova explosion,
leading to a number of competing theories of the shock-revival such as
the neutrino-driven mechanism and the magnetorotational mechanism
\citep{dimmelmeier:2008, abdikamalov:2014, Kuroda_2017, janka:2012}.

Inferring the supernova explosion (or shock-revival) mechanism has
been the primary focus of the parameter estimation literature for core
collapse GWs (see e.g., \citet{logue:2012, powell:2016, powell:2017,
  chan:2019}) and this has naturally been treated as a classification
problem due to the competing mechanisms (namely, the neutrino
mechanism and the magnetorotational mechanism) having distinct
waveform morphologies.  Other efforts have focused on estimating
various parameters that have been noted to significantly influence a
rotating core collapse GW waveform, such as the ratio of rotational
kinetic energy to gravitational potential energy of the inner core at
bounce, and the precollapse differential rotation profile
\citep{edwards:2014, abdikamalov:2014}.

The nuclear EOS, however, is a poorly understood part of physics,
though theoretical, experimental, and observational constraints are
converging, leading to greater insights about dense matter
\citep{lattimer:2012}.  It is hoped that GW detectors such as Advanced
LIGO \citep{aasi:2015}, Advanced Virgo \citep{arcernese:2014}, and
KAGRA \citep{somiya:2012} can help constrain the nuclear EOS
\citep{richers:2017}. There have been very limited attempts at
conducting parameter estimation on the nuclear EOS from rotating core
collapse GW signals. \citet{roever:2009} used a Bayesian principal
component regression model to reconstruct a rotating core collapse GW
signal and matched this to the closest waveform in the
\citet{dimmelmeier:2008} catalogue using a $\chi^2$-distance.  The EOS
of the injected signal was classified as the EOS of the best matching
catalogue signal.  The lack of success in making statistical
inferences about the nuclear EOS may perhaps be partly due to the
notion that it has very little influence on the GW signal
\citep{dimmelmeier:2008, richers:2017}.  However, in this paper, we
demonstrate that it is possible to correctly identify the nuclear EOS
at least approximately two thirds of the time.

\citet{richers:2017} provide the most in-depth study of the EOS effect
on rotating core collapse and bounce GW signal and find that the
signal is largely independent of the EOS.  However, the signal can see
stronger dependence in the post-bounce proto-neutron star (PNS)
oscillations in terms of the peak GW frequency.  They find that its
primary affect on the GW signal is through its effect on the mass of
the inner core at bounce and the central density of the post-bounce
oscillations.  We use this waveform catalogue (publicly available
through \texttt{zenodo.org} \citep{richers:data:2016}), which contains
18 different nuclear EOS, and we re-frame the problem as an 18-class
image classification and sequence classification problem, and use a
deep learning algorithm called the convolutional neural network (CNN)
to solve \citep{goodfellow:2016}.

Deep learning has already seen much success in the field of GW
astronomy.  CNNs in particular have been used for classification and
identification problems, and much of the early literature focuses on
the glitch classification problem.  For example, \citet{zevin:2017}
created the \texttt{Gravity Spy} project which uses CNNs to classify
glitches in Advanced LIGO data, with image labels outsourced to
citizen scientists.  \citet{george:2018} improve on this by using deep
transfer learning with pretrained images to get an accuracy of 98.8\%.
In terms of the GW signal identification problem, \citet{gabbard:2018}
use CNNs to identify between binary black hole signals and noise,
reproducing sensitivities achieved by matched-filtering.
\citet{george:2018b} use a CNN method called Deep Filtering to
identify binary black hole signals in noise. They also use this to
conduct parameter estimation.  Further, \citet{dreissigacker:2019} use
CNNs to search for continuous waves from unknown spinning neutron
stars.

Much effort has gone into computing low-latency Bayesian posteriors
for binary black hole systems with deep learning, particularly through
the use of variational autoencoders.  \citet{gabbard:2019} train
conditional variational autoencoders to generate Bayesian posteriors
around six orders of magnitude faster than any other method.
\citet{green:2020a} use conditional variational autoencoders in
conjunction with autoregressive normalizing flows and demonstrate
consistent results to standard Markov chain Monte Carlo (MCMC)
methods, but with near-instantaneous computation time.
\citet{green:2020b} then generalize this further to estimate
posteriors for the signal parameters of GW150914.  \citet{chua:2020}
use multilayer perceptrons to compute one and two dimensional
marginalized Bayesian posteriors.  \citet{shen:2019} use Bayesian
neural networks to constrain parameters of binary black holes before
and after merger, as well as inferring final spin and quasi-normal
frequencies.

Deep learning recently began populating the core collapse GW
literature.  \citet{astone:2018} trained phenomenological $g$-mode
models with CNNs to search for core collapse supernova GWs in
multiple terrestrial detectors.  They demonstrated that their CNN can
enhance detection efficiency and outperforms Coherent Wave Burst (cWB)
at various signal-to-noise ratios.  \citet{iess:2020} implement two
CNNs (one on time series data, and one on spectrogram data) to
classify between core collapse GW signals and noise glitches,
achieving an accuracy of $\sim 95\%$.  They also demonstrate a
proof-of-concept to classify between multiple different waveform
models, achieving an accuracy of just below $\sim 90\%$.
\citet{chan:2019} train a CNN to classify between the neutrino
explosion mechanism and magnetorotational explosion mechanism in the
time-domain. They only tested the performance of the CNN on four
signals, but achieved a true alarm probability up to $\sim 83\%$ for
magnetorotational signals at 60~kpc and up to $\sim 93$ for
neutrino-driven signals at 10~kpc, with a fixed false alarm
probability of 10\%.

In this paper, we train 2D-CNNs with 11 layers to explore visual
patterns in the rotating core collapse GW signal images, as well as a
1D-CNN with 9 layers to learn temporal patterns in the raw GW (time
series) sequence data, and make predictions about the nuclear EOS in
previously unseen test images/sequences.  The output of each network
is a vector of 18 probabilities for each image/sequence.  The EOS
class with the highest probability is the predicted EOS.  We can think
of it as the ``most likely'' EOS predicted for that GW signal.  We can
predict the EOS with up to 72\% accuracy.  If we then consider the
five most likely EOS, the signal will be correctly identified with up
to 97\% accuracy.

The paper is outlined as follows.  In Section~\ref{sec:cnn}, we
describe key elements of deep learning and discuss the CNN
architecture used in this paper.  This is followed by a description of
the data and the preprocessing required to convert it into appropriate
input images/sequences in Section~\ref{sec:preprocessing}.  We then
present results and discussion in Section~\ref{sec:results} and
concluding remarks in Section~\ref{sec:conclusions}.

\section{Deep Convolutional Neural Networks}\label{sec:cnn}

The primary objective in machine learning is to learn patterns and
rules in \textit{training} data in order to make accurate predictions
about previously unseen \textit{test} data.  \textit{Deep learning} is
an area of machine learning that transforms input data using multiple
\textit{layers} that progressively learn more meaningful
representations of the data \citep{goodfellow:2016}.  Each layer
mathematically transforms its input data into an output called a
\textit{feature map}.  The final step of each layer is to calculate
the values of the feature map using a non-linear \textit{activation
  function}.  The feature map of one layer is the input of the next
layer, allowing us to sequentially stack a network together.

One of the most popular deep learning methods, particularly in the
realm of computer vision and image classification, is the
\textit{convolutional neural network} (CNN) \citep{chollet:2018}.
Inputs into CNNs are usually 2D images, and the primary objective is
to predict the label (or class) of each image.  These are referred to
as \textit{2D-CNNs}.  Feature maps in CNNs are usually 3D tensors with
two spatial axes (height and width) and one axis that determines the
\textit{depth} of the layer.  These determine the number of trainable
parameters in each layer. Colour images (as inputs into CNNs) have
depth 3 when using the RGB colour space; one channel each for red,
green, and blue.  These can be transformed through successive layers
into feature maps with arbitrary depths, which encode more abstract
features than the three colour channels.  We can therefore think of
each layer as applying filters to its input to create a feature map.

At the final layer, we get a prediction, $\hat{y}$.  In the context of
image classification $\hat{y}$ will be a probability mass function
across all the image classes, $c = 1, 2, \ldots, C$.  This output is
compared to the truth $y$, which in image classification is a
Kronecker delta function (i.e., 1 for the true class and 0
otherwise). A distance between $y$ and $\hat{y}$ computed using a
\textit{loss function} that measures how well the algorithm has
performed when making its prediction.  The key step in deep learning
is to feed this information back through the layers in order to tune
the network's parameters.  This involves using the
\textit{backpropagation} algorithm which implements an optimization
routine to minimize the loss function, and often uses various forms of
stochastic gradient descent and the chain rule.

2D-CNNs use three different types of layers stacked together to create
a network architecture.  These are convolutional layers, pooling
layers, and fully-connected layers.  In the first instance, a
convolutional layer will apply the convolution operation to learn
abstract local patterns (such as edges) in images by considering small
2D sliding windows, producing an output feature map (of specified
depth).  Additional convolutional layers (with the previous layers'
feature map as input) then allow us to progressively learn larger
patterns in the spatial hierarchy (such as specific parts of objects)
\citep{chollet:2018}.

Pooling layers reduce the number of trainable parameters in a CNN by
aggressively downsampling feature maps, i.e., clustering neighbouring
locations of the input together using a summary statistic.  In the
case of max-pooling, the maximum value from each cluster is taken.
Pooling produces feature maps that are approximately translation
invariant to local changes in an input \citep{goodfellow:2016}.

It is often easiest to think of convolutional and pooling layers in
terms of the feature map shape (or tensor dimensions) they output,
however, fully-connected layers are best considered in terms of
neurons.  Each neuron may have many inputs $(x_1, x_2, \ldots, x_n)$
and one output $y$.  Each input has a weight $(w_1, w_2, \ldots, w_n)$
and a neuron may have bias $w_0$ associated with another input $x_0 =
1$ \citep{mackay:2003}.  The weights and bias are thought of as the
(tunable) parameters of each neuron.  The neuron is \textit{activated}
by computing the linear combination of the inputs and weights/biases
(i.e., linear activation).  It is then fed into a non-linear
activation function $f(.)$ to compute its output $y$.  That is,
\begin{eqnarray}
  a &=& \sum_{i = 0}^n w_i x_i, \\
  y &=& f(a).
\end{eqnarray}
A fully-connected layer connects one layer of neurons to another. If
there are $n$ input neurons and $m$ output neurons, the number of
tunable parameters for that layer will be $(n + 1) \times m$.

Analogous to 2D-CNNs, but for sequence processing tasks such as time
series and text sequences rather than images, are \textit{1D-CNNs}
\citep{chollet:2018}.  These function in much the same way as their 2D
counterparts, using the same three layer types.  Here, convolutional
layers consider small 1D subsequences, moving temporally, rather than
spatially, to learn local patterns in a sequence, and pooling layers
downsample reduce the length of the sequence.


Perhaps the most challenging issue with fitting CNNs is the potential
for over-fitting as there can be millions of network parameters, and
the algorithm may only memorize patterns in the \textit{training set}
and not be able to generalize these to previously unseen data
presented in the \textit{test set}.  This is why it is important to
monitor and tune a network using a \textit{validation set}.

\begin{figure}[!h]
\includegraphics[width=0.9\linewidth]{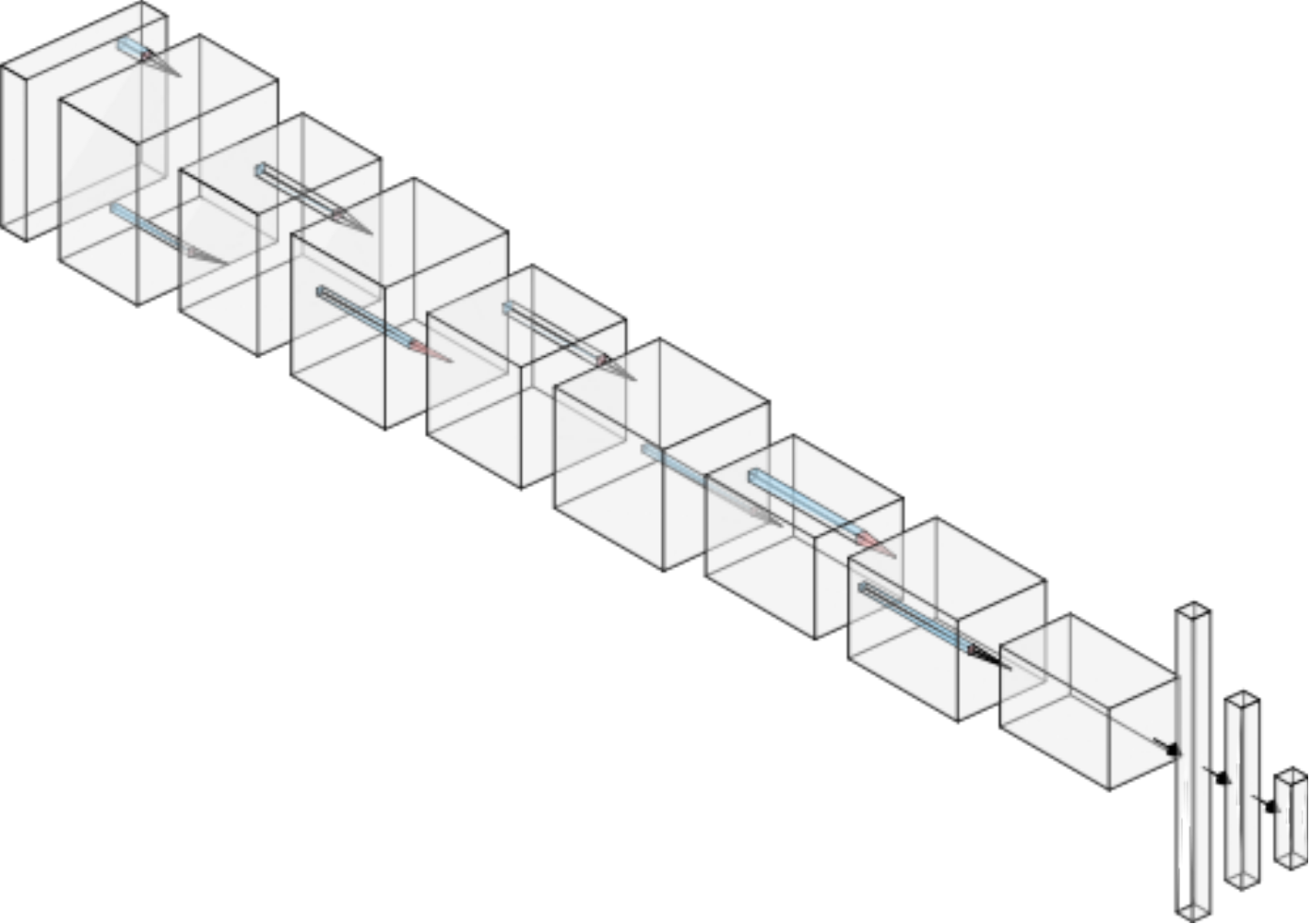}
\caption{The 2D-CNN architecture visualized.  The feature map (output)
  produced by each layer is the input into the next layer.
  Convolution and pooling layers get progressively deeper.  The height
  and width of the feature maps become smaller through pooling.}
\label{fig:architecture}
\end{figure}

In this paper, we consider both 2D and 1D variants of the CNN.  First,
we implement an 11 layer 2D-CNN.  The 11 layers of the network
architecture is outlined in Table~\ref{tab:architecture} and is
visualized in Figure~\ref{fig:architecture}.  The input layer is a 3D
tensor (image) with two spatial axes (width and height) and a depth
axis of either one (for grayscale) or three (for RGB).  Each
convolutional layer uses windows of size (3 $\times$ 3), with stride
1, and each max-pooling layer will downsample by a factor of 2.  At
the 9th layer, we ``flatten'' the output feature map from the 8th
layer to a 1D vector with the same number of neurons, which then
allows us to use fully-connected layers, connecting each neuron in the
current layer to neurons in the previous one.

For the 1D-CNN, we use a similar architecture, but with 9 layers
instead, omitting the 7th (final convolutional) and 8th (final
max-pooling) layers from the 2D counterpart for improved performance.
Similarly to the 2D-CNN, the depth of the convolutional and
max-pooling layers in the 1D-CNN sequentially increase from 32, to 64,
to 128.  Each convolutional layer uses a window length of 9 (and
stride 1), and each max-pooling layer downsamples by a factor of 4.

The choice of the number of hidden layers (and their dimensions)
depends on the data set and the task at hand, and ultimately comes
down to experimenting with different network architectures, and
monitoring the validation set error.  Though there is a lack of theory
for more than one or two hidden layers, \citet{goodfellow:2016}
demonstrated empirically that deeper networks tend to perform better
than shallower counterparts, leading to greater generalization and
higher test set accuracy.  In this study, we find that the 11 layer
2D-CNN and 9 layer 1D-CNN outlined above give us the best performance
on the stellar core collapse images and sequences respectively.  Fewer
layers tend to reduce test accuracy while additional layers add too
much complexity and increase computation time significantly.

\begin{table}[!h]
    \begin{center}
      \caption{\label{tab:architecture} The 2D-CNN architecture.  We
        use 11 layers, first sequencing between convolution and
        max-pooling layers of increasing depth. The Output Shape
        column is written as a 3D tensor with indices (Height, Width,
        Depth). We then flatten the output tensor from the 8th layer
        into a 1D vector, followed by two fully-connected layers.  It
        is easier to think of fully-connected layers in terms of the
        number of output neurons.  The final output is a probability
        mass function for the $C = 18$ different EOS classes.}
    \begin{tabular}{cccc}
    \hline
    Layer&Type&Output Shape&Activation\\
    \hline
    0&Input&(256, 256, 3)& \\
    1&Convolution&(256, 256, 32)&ReLU\\
    2&Max-Pooling&(128, 128, 32)&\\
    3&Convolution&(128, 128, 64)&ReLU\\
    4&Max-Pooling&(64, 64, 64)&\\
    5&Convolution&(64, 64, 128)&ReLU\\
    6&Max-Pooling&(32, 32, 128)&\\
    7&Convolution&(32, 32, 128)&ReLU\\
    8&Max-Pooling&(16, 16, 128)&\\
    \hline
    Layer&Type&\# Output Neurons&Activation\\
    \hline
    9&Flatten&32768&\\
    10&Fully-Connected&512&ReLU\\
    11&Fully-Connected&18&Softmax\\
    \hline
    \end{tabular}
  \end{center}
\end{table}

The rectified linear unit (ReLU) is a non-linear activation function
used on many of the layers in the network and is defined as
\begin{equation}
  f(x) = \max(0, x).
\end{equation}

The softmax function is used as the final activation, the output of
which is an 18-dimensional vector of probabilities for each
image/sequence.  This is defined as
\begin{equation}
  \hat{p}_i^{(c)} = \frac{\exp(w_c^T x)}{\sum_{c=1}^C \exp(w_c^T x)},
  \quad c = 1, 2, \ldots, C,
\end{equation}
where $x$ is the feature map from the previous layer, $w_c$ is the
vector of weights connecting the output from the previous layer to
class $c$, $C = 18$ as we have 18 different EOS we are classifying,
and $(\hat{p}_i^{(1)}, \hat{p}_i^{(2)}, \ldots, \hat{p}_i^{(C)})$ is
the vector of probabilities for the $i^{\mathrm{th}}$ image/sequence.

The loss function that we minimize is the \textit{categorical
  cross-entropy}, which is commonly-used throughout multi-class
classification problems.  This is defined as
\begin{equation}
L(p, \hat{p}) = -\sum_{i=1}^N \sum_{c=1}^C p_i^{(c)} \log\hat{p}_i^{(c)},
\end{equation}
where $N$ is the number of images/sequences in the training set and
\begin{equation}
  p_i^{(c)} =
  \begin{cases}
    1 & \text{if image/sequence $i$ belongs to class $c$}, \\
    0 & \text{otherwise}.
  \end{cases}
\end{equation}
  
We use the \texttt{RMSProp} optimizer as our gradient descent routine.
The CNN is implemented in \texttt{Python} using the \texttt{Keras}
deep learning framework \citep{chollet:2018}.

\section{Preprocessing}\label{sec:preprocessing}

We use the 1824 simulated rotating core collapse GW signals of
\citet{richers:2017}, and the data is publicly available at
\citep{richers:data:2016}.

Each signal in the data set has a source distance of 10~kpc from
Earth.  The data is originally sampled at 65535~Hz.  We downsample the
data to 4096~Hz, limiting the analysis to the most sensitive part of
the Advanced LIGO/Virgo frequency band as these ground-based GW
detectors will not be sensitive to high frequencies in the core
collapse signal due to photon shot noise.


Before downsampling, we first multiply the time-domain data by a Tukey
window with tapering parameter $\alpha = 0.1$ to mitigate spectral
leakage, and apply a low-pass Butterworth filter (with order 10 and
attenuation 0.25) to prevent aliasing.  We then downsample by removing
data according to the linear interpolation digital resampling
algorithm outlined by \citet{smith:1984}.

We align all signals such that $t_b = 0$, where $t_b$ is the time of
core bounce, and restrict our attention to the signal at times $t \in
[t_b - 0.05~\text{s}, t_b + 0.075~\text{s}]$, as this is where the
most interesting dynamics of the GW signal occur.  This is the direct
sequence input for the 1D-CNN, but further processing is required for
the 2D-CNNs.

No noise (simulated or real) is added to the signal in this paper as
our primary goal is to explore the GW signal dependence on the nuclear
EOS.

We need to produce the images to feed into the 2D-CNN.  We explore the
data in three different ways; in the time-domain with the time series
signal, in the frequency-domain with the periodogram (squared modulus
of Fourier coefficients), and in time-frequency space with a
spectrogram.

\begin{figure}[!h]
\includegraphics[width=0.9\linewidth]{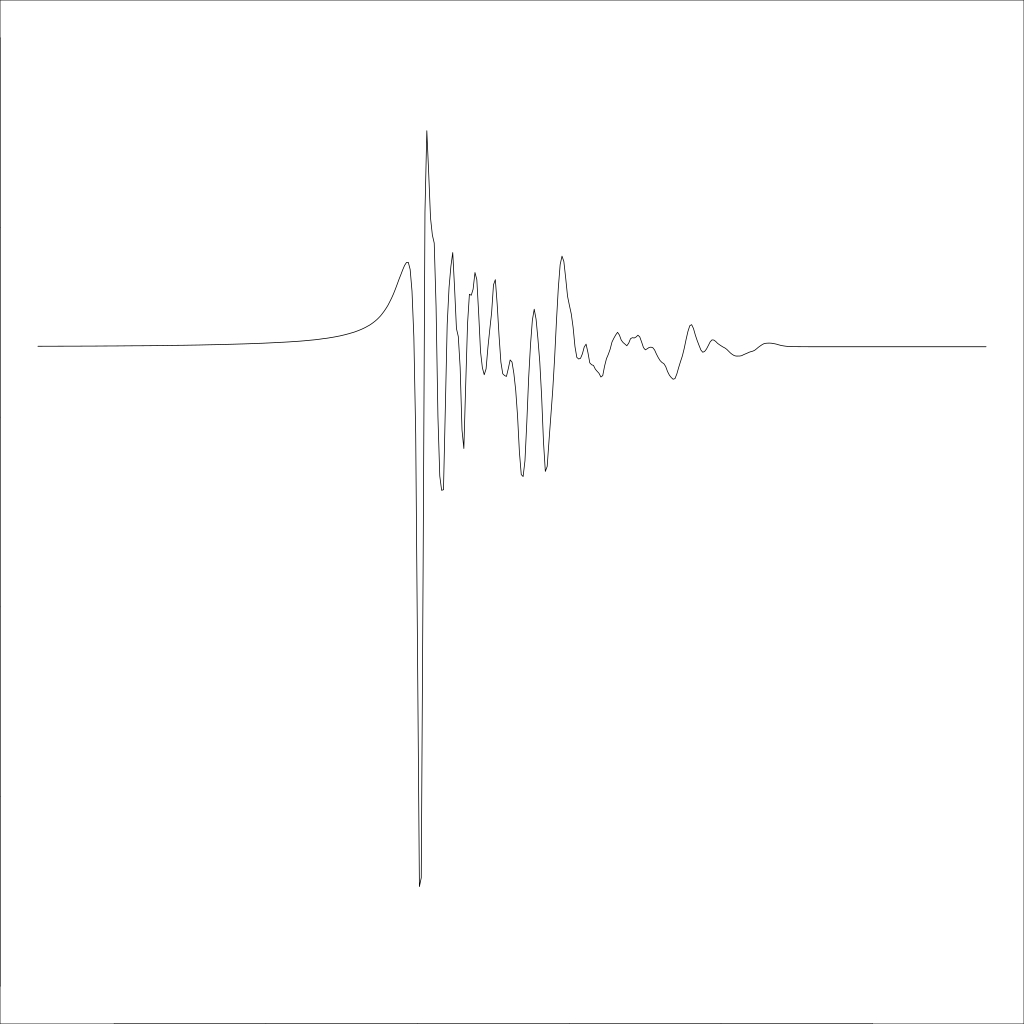}
\caption{256 $\times$ 256 pixel image of the time series of the 670th
  signal in the \citet{richers:2017} catalogue.  This signal comes
  from a $12~M_{\odot}$ progenitor, using the \texttt{HShen} EOS, with
  differential rotation of $A = 30~\mathrm{km}$, and maximum initial
  rotation rate of $\Omega_0 = 11~\mathrm{rad~s^{-1}}$.}
\label{fig:ts}
\end{figure}

First, we create images of the time-domain data.  We transform the
data set so all signals are on the unit interval.  We translate all
signals by subtracting the minimum strain from across the entire
catalogue, and then rescale by dividing by the maximum strain from
across the entire catalogue.  We plot the data, making sure to remove
the axes, scales, ticks, and labels, as these will add unwanted noise
in the image.  We then save each image as a (256 $\times$ 256) pixel
image in \texttt{jpeg} format.  An example of one of these time series
images is illustrated in Figure~\ref{fig:ts}.

\begin{figure}[!h]
\includegraphics[width=0.9\linewidth]{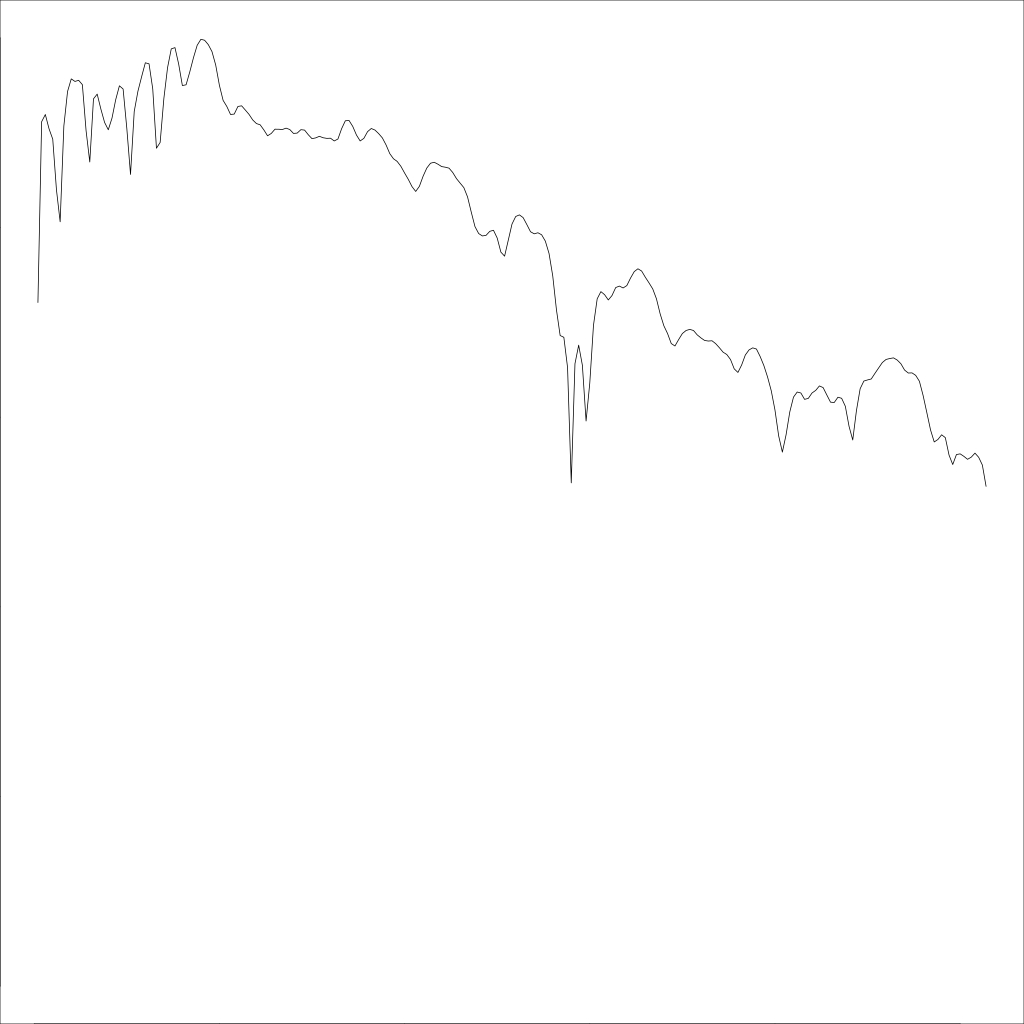}
\caption{256 $\times$ 256 pixel image of the periodogram of the 670th
  signal in the \citet{richers:2017} catalogue.  This signal comes
  from a $12~M_{\odot}$ progenitor, using the \texttt{HShen} EOS, with
  differential rotation of $A = 30~\mathrm{km}$, and maximum initial
  rotation rate of $\Omega_0 = 11~\mathrm{rad~s^{-1}}$.}
\label{fig:ft}
\end{figure}

The second set of images are the periodograms of the GW signals.  The
squared modulus of the Fourier coefficients is computed and then
transformed to the unit interval by translating and rescaling as
before (using the minimum/maximum power from the entire catalogue).
The resulting frequency-domain representations are plotted (on the
$\log_{10}$ scale) and saved in \texttt{jpeg} format as before.  The
periodogram of the signal presented in Figure~\ref{fig:ts} is
displayed in Figure~\ref{fig:ft}.

The third set of images are time-frequency maps of the data.  We
generate the ($256 \times 256$ pixel \texttt{jpeg}) images by
computing and plotting the spectrogram, which represents a signal's
power content over time and frequency.  We use a window length of
$2^7$, an overlap of 99\%, and Tukey tapering parameter $\alpha =
0.01$.

\begin{figure}[!h]
\includegraphics[width=0.9\linewidth]{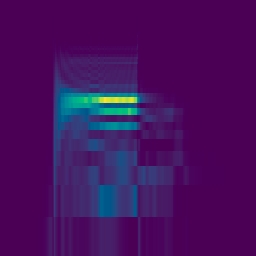}
\caption{256 $\times$ 256 image of the spectrogram of the 670th signal
  in the \citet{richers:2017} catalogue.  This signal comes from a
  $12~M_{\odot}$ progenitor, using the \texttt{HShen} EOS, with differential
  rotation of $A = 30~\mathrm{km}$, and maximum initial rotation rate
  of $\Omega_0 = 11~\mathrm{rad~s^{-1}}$.}
\label{fig:spectrogram}
\end{figure}

An example image used as input into the algorithm is presented in
Figure~\ref{fig:spectrogram}.  Note that the frequency axis is on the
$\log_2$ scale, and power (colour) is normalized by dividing the power
in each of the spectrograms by the maximum total power from across the
entire catalogue to ensure images are all on the same scale.  As
before, axes, ticks, scales, and labels are removed.

For each of the three image data sets and the one sequence data set,
we then randomly shuffle the images/sequences such that $\sim 70\%$
are in the training set ($n_{\mathrm{training}} = 1302$), $\sim 15\%$
are in the validation set ($n_{\mathrm{validation}} = 261$), and $\sim
15\%$ are in the test set ($n_{\mathrm{test}} = 261$).

We run three separate 2D-CNNs (one each for the time series images,
periodogram images, and spectrogram images) to explore visual
patterns, and one 1D-CNN on the sequence data to explore temporal
patterns, with the goal of classifying nuclear EOS.

The input depth for the time series and periodogram images is one
grayscale colour channel, whereas for the spectrogram images, this is
a three colour RGB channel.  The input depth for the sequence data is
one as we only have univariate time series data.

\section{Results}\label{sec:results}

We measure the success of the four CNNs in terms of the proportion of
test signals that have the correct EOS classification, called the
\textit{accuracy} of the network.  In this study, we achieve 64\%
accuracy for the spectrogram images, 65\% for the periodogram images,
71\% for the time series images, and 72\% for the direct sequence
data.

State-of-the-art CNNs can achieve accuracies of up to 95-99\% on
every-day objects in computer vision competitions such as those based
on the ImageNet database \citep{imagenet:2015}.  This has been
demonstrated effectively in the GW literature (see e.g.,
\citep{george:2018}).  Though our achieved accuracy of 64--72\% is
lower than this, it is much higher than anticipated.  As noted by
\citet{richers:2017}, the rotating core collapse GW signal has only
very weak dependence on nuclear EOS.  Our results suggest that this
could be upgraded to ``moderate'' dependence.  What is also surprising
is the algorithm achieved this accuracy with a relatively small
training data set ($n = 1302$).

Let us now consider the ``top 5'' EOS classifications for each
image/sequence.  That is, the five EOS classes with the highest
probabilities for each image/sequence.  We compute the cumulative
proportion of images/sequences in the test set that are correctly
classified within these top 5 classes.  The cumulative proportion of
correct classifications can be seen in Table~\ref{tab:cumulative}.
Interestingly, the 2D-CNN trained on time series images outperforms
the other 2D-CNNs, and the 1D-CNN does slightly better than this.  For
each CNN, the EOS class with the second highest probability is the
correct classification on more than 10\% of the test signals,
indicating that we can correctly classify the EOS within the top 2
classes 75--88\% of the time.  For the 1D-CNN and the time series
2D-CNN, we achieve more than 90\% correct classifications within the
top 3 EOS classes.  We can can correctly constrain the nuclear EOS to
one in five classes (rather than one in 18) 97\%, 93\%, 91\%, and 97\%
of the time for the time series 2D-CNN, periodogram 2D-CNN,
spectrogram 2D-CNN, and sequence 1D-CNN respectively.  These results
are encouraging and demonstrate that we can constrain the nuclear EOS
with reasonable accuracy.  It is worth noting for the spectrogram
images that a 2D-CNN with one grayscale input colour channel yields
consistent results to the 2D-CNN with three RGB input colour channels
presented here.  However, these results have been omitted for brevity.

\begin{table}[!h]
    \begin{center}
      \caption{\label{tab:cumulative} Cumulative proportion of correct
        classifications.}
    \begin{tabular}{lcccc}
      \hline
      &\multicolumn{3}{c}{2D-CNN} & \multicolumn{1}{c}{1D-CNN} \\
    &Time Series&Periodogram&Spectrogram&Sequence \\
    \hline
    1&0.71&0.65&0.64&0.72 \\
    2&0.85&0.77&0.75&0.88 \\
    3&0.91&0.85&0.83&0.91 \\
    4&0.93&0.90&0.85&0.94 \\
    5&0.97&0.93&0.91&0.97 \\
    \hline
    \end{tabular}
  \end{center}
\end{table}

Although \citet{iess:2020} found that their 2D-CNN on spectrogram
images slightly outperformed their 1D-CNN on sequence data (due to
common features in the spectrograms), we find the opposite here.  The
raw GW sequences are the purest form of the data.  This is
particularly true when no noise is added to the signal, as assumed
here.  Converting time series to images requires further preprocessing
with certain user decisions to be made.  This could create
image-induced uncertainty, which could have an effect on the
predictive power of the 2D-CNNs.  For example, spectrogram images are
subject to choices in window type, window length, and overlap
percentage, as well as plotting decisions such as image resolution,
and results could depend on these choices.  Therefore, the superior
accuracy of the 1D-CNN is not surprising in the present work.

We ran the CNNs in batches of size 32 for 100 epochs for the 2D-CNNs
and 30 epochs for the 1D-CNN, making sure to monitor validation
accuracy and loss.  Overfitting was not an issue with the 2D-CNNs,
even though it is a relatively small data set.  No regularization,
drop-out, or $K$-fold validation was required.  While training
accuracy tended towards 100\% as the number of epochs increased,
validation accuracy remained reasonably constant at 60--70\% after
about 40 epochs for the 2D-CNNs, and this translated to the test set.
Validation loss did not noticeably increase as the epochs increased.
The 1D-CNN required fewer epochs and started noticeably overfitting
after about 30 epochs.

\begin{figure}[!h]
\includegraphics[width=0.9\linewidth]{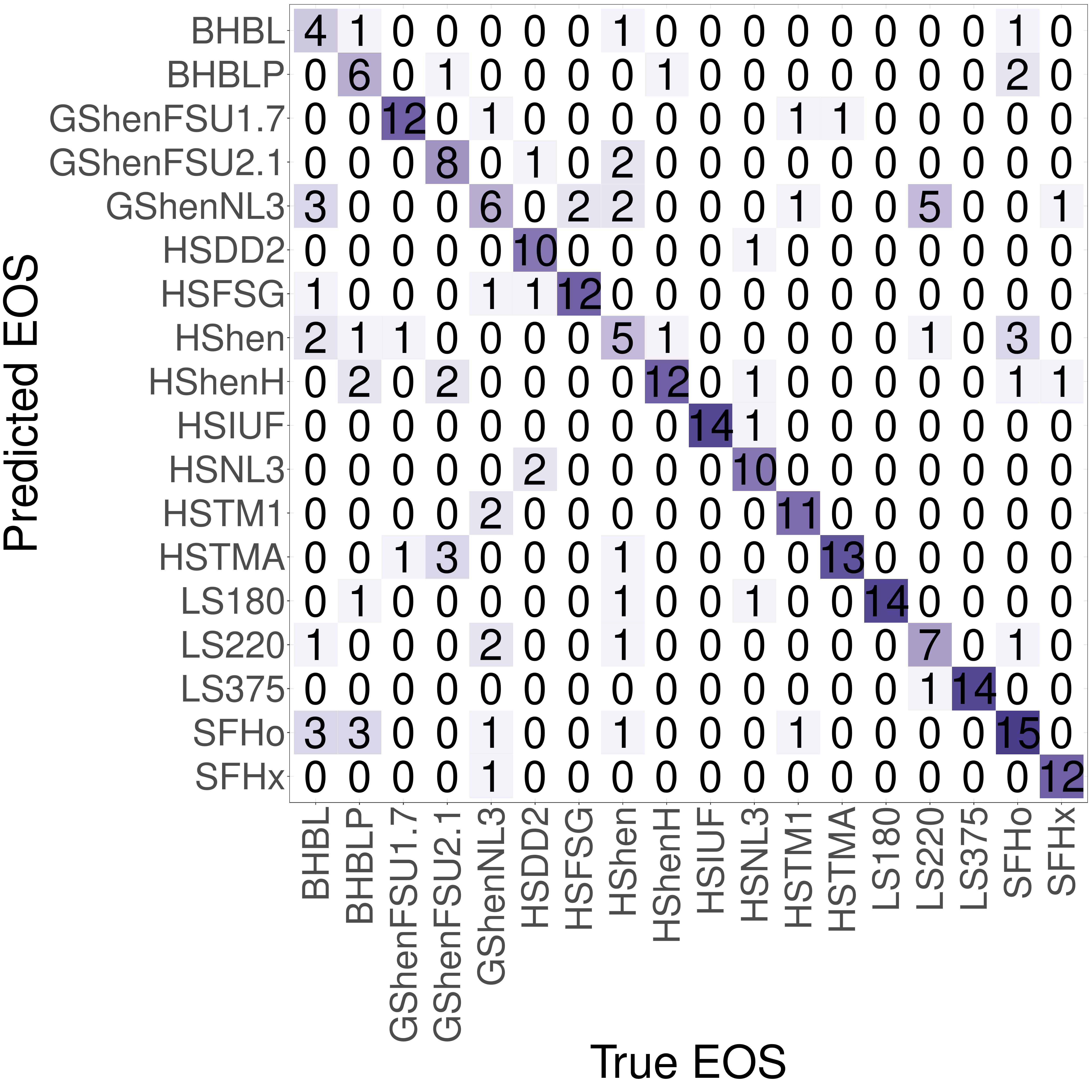}
\caption{Confusion matrix for the test set of time series images.  The
  darker the colour, the higher the number of correct
  classifications.}
\label{fig:cm}
\end{figure}

In Figure~\ref{fig:cm}, we produce a confusion matrix that compares
the true EOS class against the predicted EOS class for the test set of
time series images.  The confusion matrix gives us information on
which EOS classes are well-classified, and which ones the CNN
struggles to classify.  For example, we can see that the CNN can
classify the \texttt{GShenFSU1.7}, \texttt{HSDD2}, \texttt{HSFSG},
\texttt{HShenH}, \texttt{HSIUF}, \texttt{HSNL3}, \texttt{HSTM1},
\texttt{HSTMA}, \texttt{LS180}, \texttt{LS375}, and \texttt{SFHx} EOS
with very good accuracy (at least 10 out of 14 signals), and
\texttt{SFHo} with moderate accuracy of 15 out of 23 signals.  We
also see that the \texttt{BHBL} and \texttt{HShen} EOS are relatively
poorly classified, noting that the \texttt{BHBL} EOS is often confused
as the \texttt{GShenNL3}, \texttt{HShen}, and \texttt{SFHo} EOS, and
the \texttt{HShen} EOS confusion is spread amongst many different EOS.

As discussed by \citet{richers:2017}, the EOS that are in best
agreement with experimental and astrophysical constraints are
\texttt{LS220}, \texttt{GShenFSU2.1}, \texttt{HSDD2}, \texttt{SFHo},
\texttt{SFHx}, and \texttt{BHBLP}.  From the confusion matrix, we see
that of these more astrophysically realistic EOS, \texttt{HSDD2},
\texttt{SFHo}, and \texttt{SFHx} are classified well, with little
confusion.  However, the other three astrophysically realistic EOS
classes are misclassified $\sim 50\%$ of the time.  Of note, the
\texttt{LS220} EOS is misclassified as the \texttt{GShenNL3} EOS for 5
out of the 14 test signals in that class.  We can also see that the
\texttt{BHBLP} EOS often gets confused with the \texttt{SFHo} and
\texttt{HShenH} EOS, and the \texttt{GShenFSU2.1} EOS is often
confused with the \texttt{HSTMA} and \texttt{HShenH} EOS.



\section{Conclusions}\label{sec:conclusions}

This paper demonstrated a proof-of-concept that rotating core collapse
GW signals moderately depend on the nuclear EOS.  We are encouraged by
the 64--72\% correct classifications achieved when using the CNN
framework to probe visual and temporal patterns in rotating core
collapse GW signals.  We are further encouraged by the 91--97\%
correct classifications after considering the five EOS classes with
the highest estimated probability for each test signal.  With this in
mind, we plan a follow-up study to explore further how the feature
maps of each layer can help understand exactly how each nuclear EOS
influences the GW signal.

The goal of this paper was not to conduct parameter estimation in the
presence of noise, but more to explore the dependence a rotating core
collapse GW signal has on the nuclear EOS.  However, this is a goal of
a future project, where we aim to add real or simulated detector noise
to see if we can constrain nuclear EOS under more realistic settings.

The deep learning framework is becoming a force of its own in the GW
data analysis literature; allowing for near-instantaneous low-latency
Bayesian posterior computations using pre-trained networks, producing
accurate and efficient GW signal and glitch classifications, and
allowing us to solve problems previously thought impossible.

\section*{Acknowledgements}

The author would like to thank Nelson Christensen, Ollie Burke, and
Hajar Sadek for fruitful discussions.


\bibliographystyle{spbasic}
\bibliography{refs}  

\end{document}